\documentclass{article}
\usepackage{sao2}
\usepackage{psfig}
\issue{1999, 47, 5-185}
\begin{document}
\markboth{Karachentsev et al.}{The Revised Flat Galaxy Catalogue}
\title{The Revised Flat Galaxy Catalogue}
\author{I.D. Karachentsev \inst{a} \and V.E. Karachentseva \inst{b}
\and Yu.N. Kudrya \inst{b} \and M.E. Sharina \inst{a}
\and S.L. Parnovsky \inst{b}}
\institute{Special  Astrophysical Observatory of the Russian AS, Nizhnij
Arkhyz 357147, Russia
\and  Astronomical Observatory of Kiev University, Observatorna 3,
   Kiev 254053, Ukraine}
\date{May 5, 1999}{May 14, 1999}
\maketitle
\begin{abstract}  We present a new improved and completed version of the
Flat Galaxy Catalogue (FGC) named the Revised Flat Galaxy Catalogue (RFGC)
containing 4236 thin edge-on spiral galaxies and covering the whole
sky. The Catalogue is intended to study large-scale cosmic streamings
as well as other problems of observational cosmology.
The dipole moment
of distribution of the RFGC galaxies ($l = 273\degr,\; b =+19\degr$) lies within
statistical errors ($\pm10\degr$) in the direction of the Local Group
motion towards the Microwave Background Radiation (MBR).

\end{abstract}
\section{Introduction}

   The Catalogue of flat spiral edge-on galaxies FGC (Karachentsev et
al., 1993a) represents a rather specific sample of 4455 galaxies
satisfying two simple conditions:
\begin{itemize}
 \item   the axial ratio for the blue image $a/b\geq7$;

  \item the angular diameter (blue major axis of galaxy) corresponds to $a \geq0\farcm6$.
\end{itemize}
   Due to this selection criterion, the Catalogue is morphologically
homogeneous and contains more than 75\% of Sc and later spiral types.

  As argued by Karachentsev (1989), such  thin edge-on spirals are
an appropriate tool to study the large-scale motions in the Universe
because of : a) the HI 21 cm  and $\rm H_\alpha$ line detection rate of these
galaxies is nearly 100\%; b) the flat galaxies avoid volumes occupied by
groups and clusters so that their structure remains undisturbed and they
are not affected by large virial motions.

   Selection of objects and determination of their characteristics were
carried out by systematic visual inspection of all prints of the
Palomar Observatory Sky Survey (POSS-I) and the ESO/SERC sky survey
in the blue and red colours. In accordance with the original photografic
material, the Catalogue consists of two parts: FGC (N=2573) and its southern
extension, FGCE (N=1882). The first part is based on the POSS-I and covers the
sky region with declinations between $-20\degr$ and $+90\degr$. The second
one is based on the ESO/SERC and covers the rest of the sky area up to
the DEC=$-90\degr$.
   Besides, 291 galaxies selected in the preliminary survey were rejected
then from the main Catalogue due to the violation of the  $a/b \geq7$ criterion.
They were included in Addendum.

   General properties of flat galaxies have been studied in detail in a
series of our papers (Karachentsev et al., 1993b; 1996; 1997a; 1998;
Kudrya et al., 1994; 1997a,b; Parnovsky et al., 1994; Karachentsev, 1999).

   To obtain the distances of flat galaxies independently of their
radial velocities, extensive HI observations with the 305-m
radiotelescope of the Arecibo observatory (Giovanelli et al., 1997) as
well as the $V_{\rm rot}$ observations with the SAO 6 m  telescope (Makarov et
al., 1997a,b) have been performed. These observations added by the
literature data allowed   peculiar velocities for about
900 flat galaxies to be calculated and the apex of their coherent motion
relative to the MBR frame to be determined (Karachentsev et al., 1995).

   At present, the FGC is the deepest, morphologically
homogeneous and  complete sample of field galaxies suitable for
investigation   of different problems of extragalactic
astronomy and observational cosmology.

\section{New version of FGC}

   The main reasons for the preparation of a new improved and supplemented
Catalogue version were as follows:
\begin{itemize}

\item  possibility of remeasuring the coordinates of flat galaxies with
an accuracy  $\simeq3\arcsec$, higher than in the FGC,  using the Digital Sky Survey;

\item   inclusion of data about ``red'' galaxy diameters lacking in the FGC;

\item  reduction of the diameters measured on the J and R films of the
ESO/SERC to the diameter system of the POSS-I (close to $a_{25}$ system),
which decreased the difference in photometric depth between the two parts
of the Catalogue (Kudrya et al., 1997a);

\item   calculation of total apparent magnitudes (with a standard error
 $\sim$ 0.25 mag) for all flat galaxies, basing on angular diameters,
surface brightnesses, and other parameters (Kudrya et al., 1997b);

 \item   possibility of determining the values of galactic extinction  towards
 each flat galaxy,  using the new IR data (Schlegel et al., 1998).

 \item   necessity for removing some faults noticed during the work with the FGC
data.
\end{itemize}

   The differences in structure  between the new and old  Catalogue versions are
the following:
\begin{itemize}
\item both parts, FGC and FGCE, have been joined in the RFGC (Revised
Flat Galaxy Catalogue) where the galaxies are ranged according to their
right ascensions for the epoch J2000.0;

\item  the Addendum has been omitted;

\item the Notes describing specific galaxy  characteristics have been included
in the main body of the Catalogue (some details omitted);

\item the lists of identification of the FGC and the FGCE galaxies have
been omitted because these data are now accessible  from different
galaxy databases (NED, LEDA etc).
\end{itemize}

    As a result, the RFGC Catalogue contains the following data:

    {\bf column 1:} \phantom{a} new (RFGC) galaxy number. The galaxies having the
reduced diameters,
$a{\rm_O}$, less than 0.6 arcmin are left in the Catalogue without the new RFGC
number;

    {\bf column 2:} \phantom{a} old Catalogue number; the letter ``E'' is added for FGCE galaxies;

    {\bf column 3:} \phantom{a} PGC (Paturel et al., 1989) galaxy number;

    {\bf columns 4,5:} \phantom{a} Right Ascension and Declination for the epoch J2000.0;

    {\bf columns 6,7:} \phantom{a} Right Ascension and Declination for the epoch J1950.0;

    {\bf columns 8,9:}  \phantom{a} galactic longitude and latitude determined from RA, DEC
with the North Galactic pole at RA=$12^h49^m$ and DEC=$+27\fdg4$;

    {\bf column 10:} \phantom{a} positional angle of the galaxy major axis measured  north ---
east in deg.;

    {\bf columns 11,12:}\phantom{a} $a{\rm_O},\;b{\rm_O}$ --- major and minor blue diameters in arcmin in the POSS-I
diameter system. For FGCE galaxies a conversion from the ESO/SERC
diameter system (J) to the POSS-I one (O) was done according to
the relations: $a{\rm_O} = 0.8078 a_J,\; b{\rm_O} = 0.7827 b_J$;

    {\bf columns 13,14:}\phantom{a} $a{\rm_E},\; b{\rm_E}$ --- major and minor red diameters in arcmin
in the POSS-I diameter system.  The conversion
formulae for the FGCE galaxies are $a{\rm_E} = 0.8640 a_R,\; b{\rm_E} = 0.9730 b_R$;

   {\bf column 15:}\phantom{a} $B_t$ --- total apparent magnitude calculated using the data
on blue angular diameters, morphological type, surface brightness,
and ``colour index'', $\log a{\rm_O}/a_{\rm_E}$, according to the relation:

$$B_t = \left\{
 \begin{array}{ll}
1.22x^c+14.89, & x^c\leq-0.9 \\
0.78x^c+14.44, & x^c>-0.9\\
\end{array} \right \} + $$
$$+0.15(SB-1.9)-0.023(T-5.4)+1.4\log(a{\rm_O}/a{\rm_E}), $$
where $$x^c   = -2.5 \log (a{\rm_O}\cdot b{\rm_O}) + 0.05\log (a{\rm_O}/b{\rm_O}),$$ and
SB is the surface brightness index, T --- galaxy type  (see below);

    {\bf column 16:}\phantom{a}  value of galactic extinction in the B band;

    {\bf column 17:}\phantom{a} morphological type of the spiral according to the Hubble
classification. Note that Sb = 3, Sc = 5 etc;

    {\bf column 18:}\phantom{a} index of apparent asymmetry of the galaxy shape (0 means
poorly defined, 2 --- pronounced);

    {\bf column 19:}\phantom{a} index of the mean surface brightness (I --- high, IV --- very
low);

    {\bf column 20:}\phantom{a} number of significant neighbours with an angular
diameter in the range from $a{\rm_O}/2$ to $2a{\rm_O}$ in a circle of $R=10 a{\rm_O}$,
where $a{\rm_O}$ is the blue major axis of the galaxy considered;

    {\bf column 21:}\phantom{a} notes describing the galaxy morphological peculiarities
and/or the galaxy environment. The galaxy
diameters and mutual distances are expressed in arcmin.

\section{Some statistics and data description}

   The distribution of the numbers of the RFGC galaxies according to their
Catalogue characteristics are presented in Tables 1--11.

   The galaxies were divided into four angular diameter intervals: with
$a{\rm_O} \geq 2\farcm0,\;  1\farcm5\leq a{\rm_O}\leq 1\farcm99,\;
1\farcm0 \leq a{\rm_O} \leq 1\farcm49$, and $0\farcm6\leq a{\rm_O}\leq 0\farcm99$.
Based on the  radial velocity and
inverse angular diameter relation for about 1000 RFGC galaxies, the
 mean distances, in km/s, are equal to 3860, 5150, 7730, 12900
for $a{\rm_O} = 2\farcm0,\; 1\farcm5,\; 1\farcm0$, and $0\farcm6$, respectively.

   We picked out also the galaxies of different flatness: from the axial
ratio being near our selection limit, $7.0 \leq a/b \leq 7.99$, to the flattest
ones, with  $a/b \geq 10$.

   The results presented in the tables are clear and do not need special explanations.
   Comment briefly on some  selection effects.
\begin{itemize}
\item  Near the limit  of the RFGC angular diameter a deficit of very flat
galaxies is seen --- about 6\% in comparison with that expected  for a
random
distribution (Table 1). This selection may be caused by the emulsion
resolution effect. For  example, a galaxy with  $a=36\arcsec$ and $a/b=10$
has a  minor axis value,
$b = 3.6\arcsec$, comparable with a typical 2$\arcsec$ seeing.

   \item The numbers of different galaxy types in different intervals according
to their angular diameters  are close to expected. Only small Sb type
excess is found among large galaxies (Table 3).

   \item  A small deficit of high surface brightness galaxies among
the smallest ones can be explained as an  effect of  finite resolution
of the emulsion (Table\,4).

   \item  The largest galaxies seem to be more isolated (a small excess of galaxies
without significant neighbours) (Table 6). A similar effect
was noted for isolated galaxies (Karachentseva, 1973).

   \item  The flattest galaxies demonstrate  excess of
very low surface brightness galaxies (Table 9), as well as excess
of isolated ones (Table 11).
\end{itemize}

   The integral distribution of the RFGC galaxies on their blue major
diameters, $\log N$ vs. $\log a$, is presented in Fig.1. The slope of the linear part
is equal to 2.50 (for red diameters the slope is 2.53).
 As it is seen from Fig.1, the RFGC is complete
to $a{\rm_O}=0\farcm9$. There are no apparent signs of the Local Supercluster
presence.

   Fig.2 shows the integral distribution of the RFGC galaxies
on their blue axial ratios, $\log N$  vs. $a/b$. In the interval of $a/b$ from 7 to 19
the relation is exponential. The maximum axial ratio  equals  21 (for
red diameters $(a/b)_{\rm max}$ is equal to 19).

   The all-sky distributions of the RFGC galaxies in the equatorial,
galactic, and supergalactic coordinates are displayed in Fig.3 a--c.
It is  rather homogeneous, without a distinct difference in galaxy density between
two parts of the sky, above and below DEC=$-18\degr$. The RFGC galaxies do
not exhibit density concentration towards the supergalactic equator
in accordance with the data of Fig.1.

   Comparing the  galaxy distributions on their positional angles at
different distances (Fig.4 a--d), we see that the anisotropy in these
distributions increases from nearby $(a\geq2\arcmin)$ to  distant RGFC galaxies.

   Fig. 5 a,b  shows the distribution of the normals to flat galaxies
using an equal-area projection in galactic coordinates
$r=\sin(\pi/4-b/2),  \phi=l$, where  the senses of the
northern hemisphere have been chosen for each galaxy. The central void is produced
by the Zone of Avoidance along the galactic equator. All RFGC galaxies exhibit an increase
of pole concentration in a wide region around $(l =60\degr,\; b=+25\degr)$, but for the nearest
galaxies their pole distribution looks homogeneous. Hence, the positional angle
anisotropy has a scale larger than that of the Local Supercluster.

    The distribution of the RFGC galaxies in galactic coordinates
vs.  their diameter (distance) intervals is exhibited in Fig.6 a--d.

   The nearest galaxies are seen to be slightly concentrated in
the Virgo region,  while the most distant galaxies are concentrated in the region of the
Hydra-Pavo-Indus supercluster. The RFGC galaxies lying at moderate distances
are distributed  homogeneously over the sky.

   The distribution of the RFGC galaxies over the sky is different for
 galaxies of different flatness, Fig.7 a--c. The galaxies having
$a/b$ near the selection limit are distributed quite homogeneously,
while the galaxies with  moderate axial ratios show a slight concentration towards the
Hydra cluster region. The flattest galaxies show a pronounced clustering
in accordance with our result obtained from  the two-point angular correlation
function for flat galaxies (Karachentsev et al., 1996).

\section{Optical dipole}
  As it is well known, the dipole anisotropy of the
MBR points out that the Local Group moves at a velocity of
600 km/s with
respect to the MBR towards the apex with
galactic coordinates $l= 268\degr,\; b= +27\degr$
(Kogut et al., 1993). The way to study the cause of this motion is
to study the dipole moment of distribution of galaxies in the all-sky
catalogue. This approach has been applied by Harmon et al. (1987) to
the ``IRAS Point Source Catalog'' and by Lahav et al. (1988) to the combination
of three optical catalogues: UGC+ESO+MCG (Nilson, 1973; Lauberts, 1982;
Vorontsov-Velyaminov et al., 1962--64). According to these authors
the resulting IRAS and optical ``UGC+ESO+MCG'' dipoles lie within about $10\degr$
of the MBR dipole.

   Based on the coordinates of the RFGC galaxies in different angular diameter
intervals, we calculated their centroid position, when the galaxies were
equally (not by flux) weighed. The results are presented in Table 12.
Here  X,Y, and Z indicate the average galactic Cartesian coordinates
for the galaxy centroid together with their standard errors, and the
corresponding ${l,b}$ coordinates show where the optical dipole vectors
have a sense of  the sky. As it is seen from these data the whole  sample
dipole ($l=273\degr,\; b=+19\degr$) lies within $10\degr$ of the MBR dipole.
This alignment may indicate that the Local Group motion with respect to
the MBR is produced mainly by  matter distribution on a scale of
$\sim$15000 km/s sampled by the RFGC.

\begin{acknowledgements}
   For measuring the coordinates we  used the Digitized
Sky Survey. The Digitized Sky Survey was produced at the Space
Telescope Science Institute under U.S. Government grant NAG W-2166.
The authors are greatly indepted to G.G. Korotkova for technical assistance
in preparing the Catalogue.
   This work was supported by the grant of RFBR 98-02-16100.
\end{acknowledgements}

{}

\begin{figure*}
\centerline{\psfig{figure=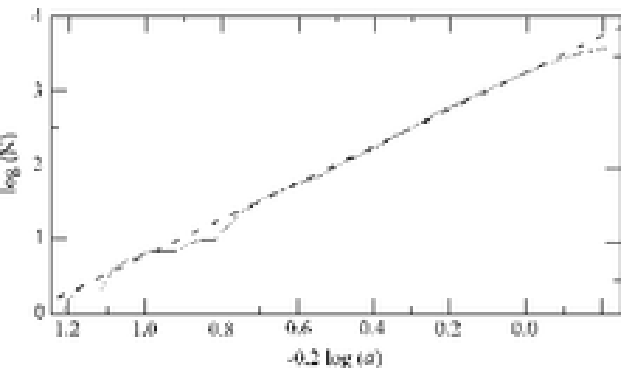,width=16cm}}
\caption{The integral distribution of the RFGC galaxies on their blue major
angular diameter.}
\end{figure*}
\begin{figure*}
\centerline{\psfig{figure=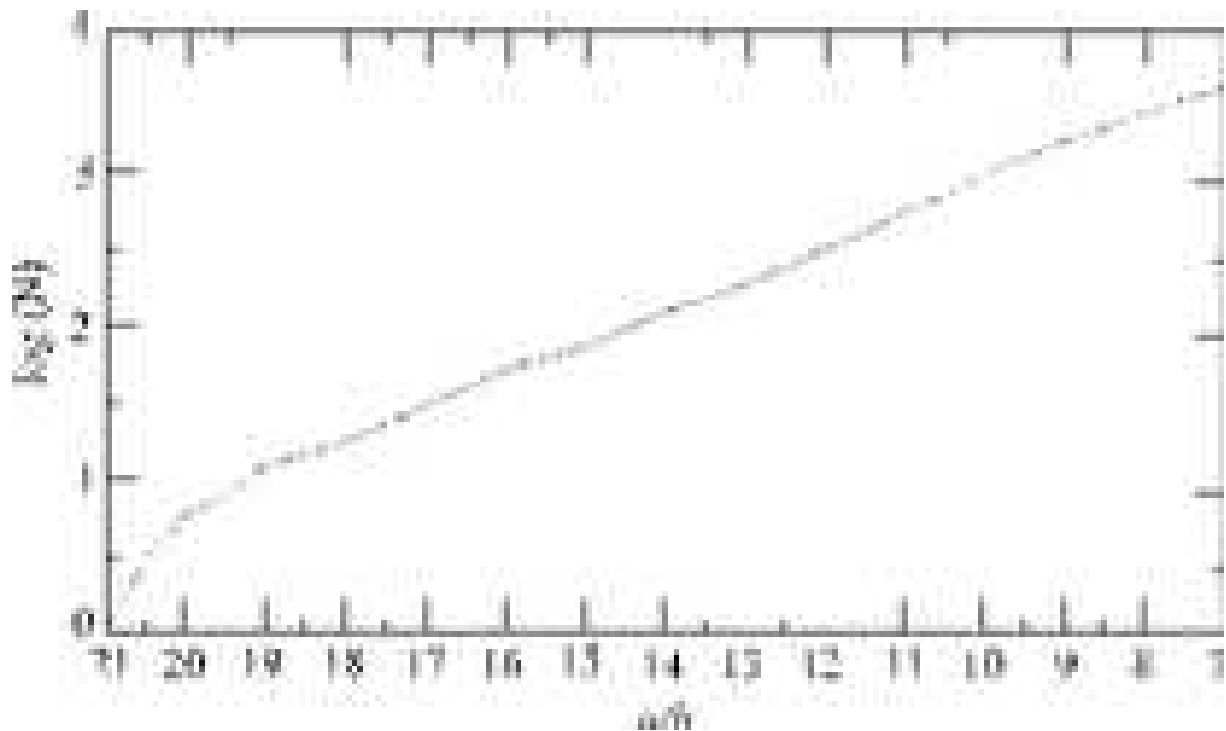,width=16cm}}
\caption{The integral distribution of the RFGC galaxies on their blue
axis ratios.}
\end{figure*}

\begin{figure*}
\centerline{\psfig{figure=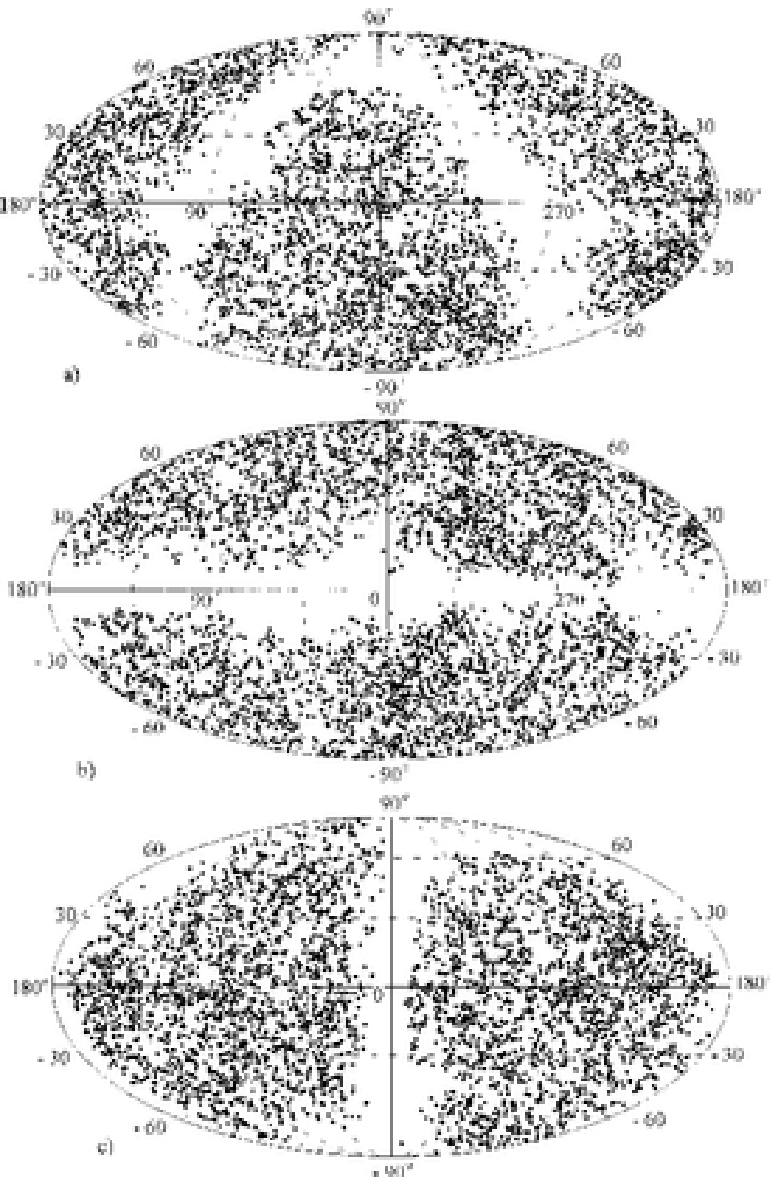,width=13cm}}
\caption{The whole sky distribution of the RFGC galaxies: \newline
a) in equatorial coordinates, b) in galactic coordinates, c) in supergalactic
coordinates.}
\end{figure*}

\begin{figure*}
\centerline{\psfig{figure=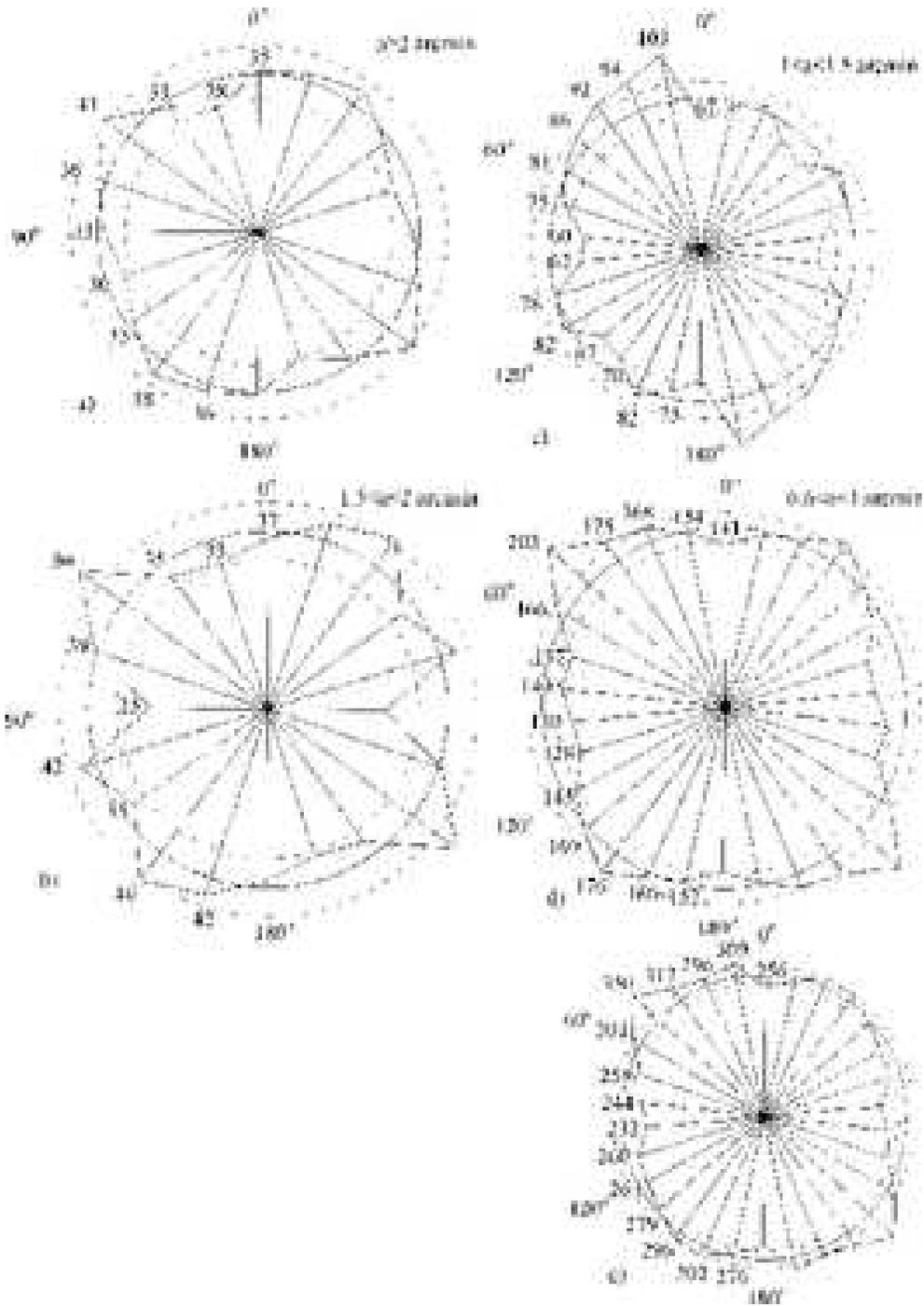,width=13cm}}
\caption{The distribution of the RFGC galaxies on their positional angles.
The numbers of galaxies in each PA sector are marked. Two dashed lines indicate
$\pm \sigma$ deviation from the average number of galaxies indicated by
solid line: \newline
a) $a>2$ arcmin, b) 1.5 arcmin$ <a<2$ arcmin, c) 1.0 arcmin$ <a<1.5$ arcmin,
\newline d) 0.6 arcmin$ <a<1.0$ arcmin, e) all galaxies.}
\end{figure*}

\begin{figure*}
\centerline{\psfig{figure=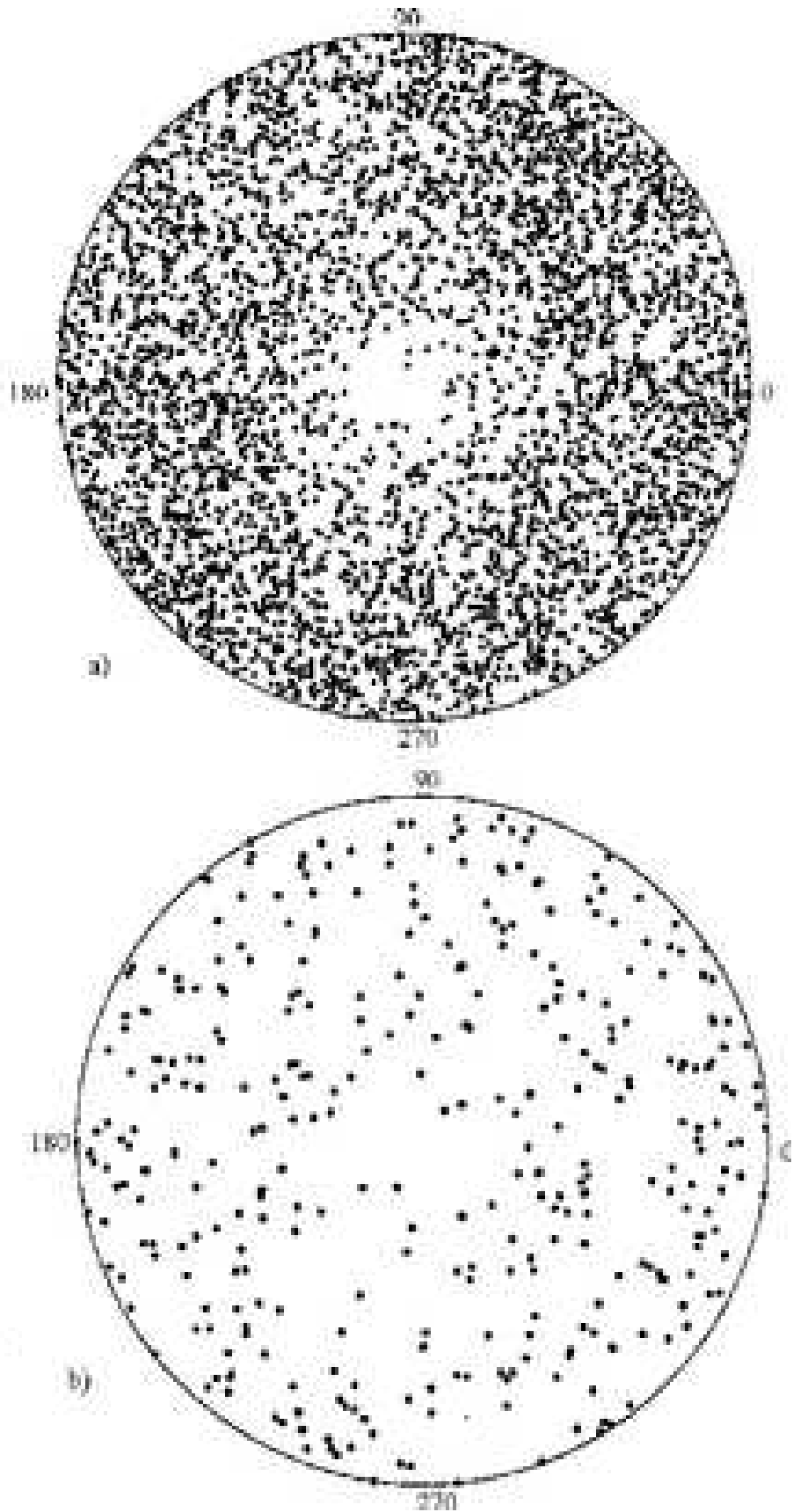,width=12cm}}
\caption{The distribution of normals to the RFGC galaxies over the sky:
\newline a) all galaxies, b) the galaxies with $a>2$ arcmin.}
\end{figure*}

\begin{figure*}
\centerline{\psfig{figure=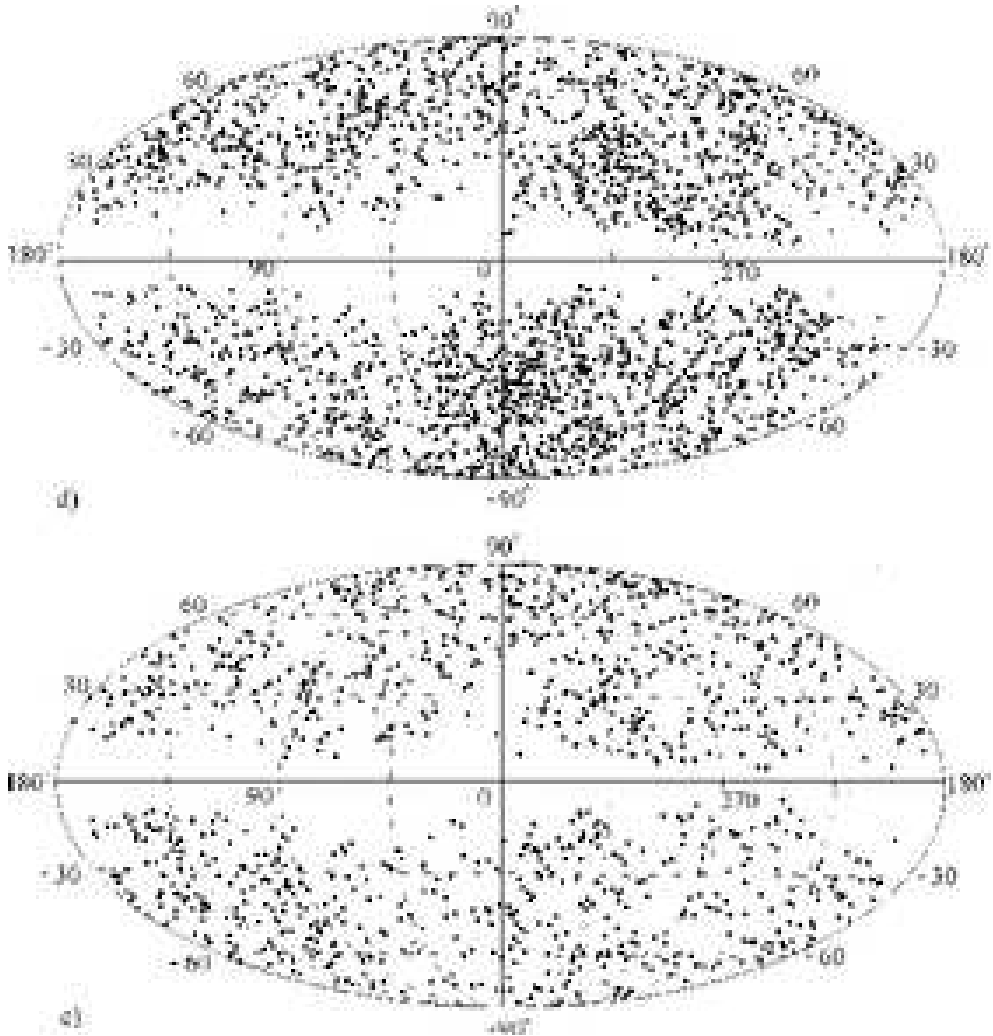,width=13cm}}
\caption{a),b). The distribution of the RFGC galaxies over the sky in galactic
coordinates: \newline a) $a>2$ arcmin, b) 1.5 arcmin$ <a<2$ arcmin.}
\end{figure*}

\begin{figure*}
\setcounter{figure}{5}
\centerline{\psfig{figure=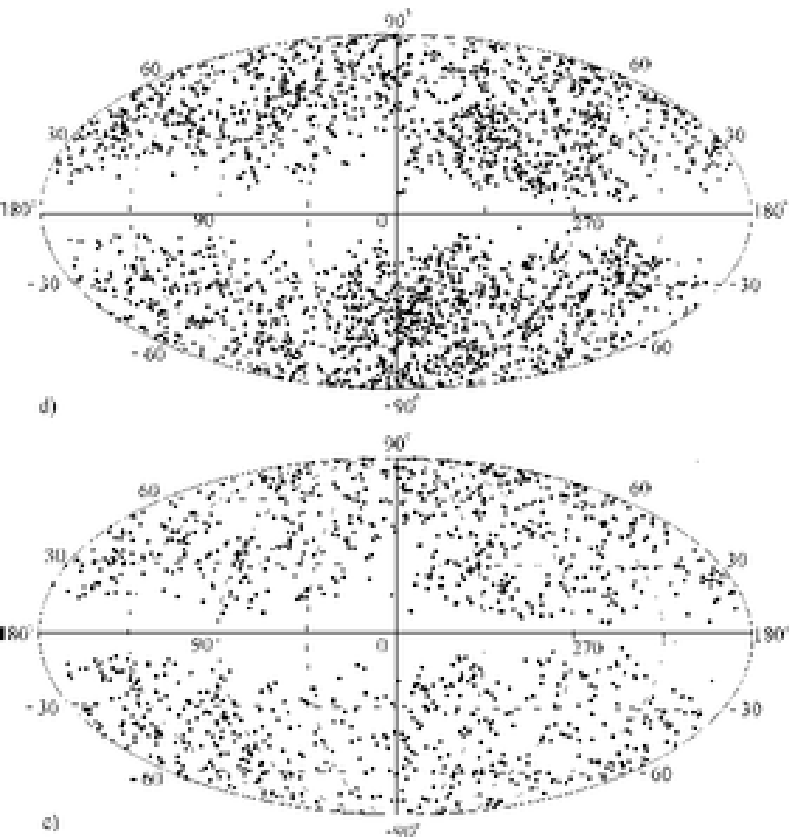,width=13cm}}
\caption{c),d) The distribution of the RFGC galaxies over the sky in galactic
coordinates: \newline c) 1.0 arcmin$ <a<1.5$ arcmin, d) 0.6 arcmin$ <a<1.0$ arcmin.}
\end{figure*}

\begin{figure*}
\centerline{\psfig{figure=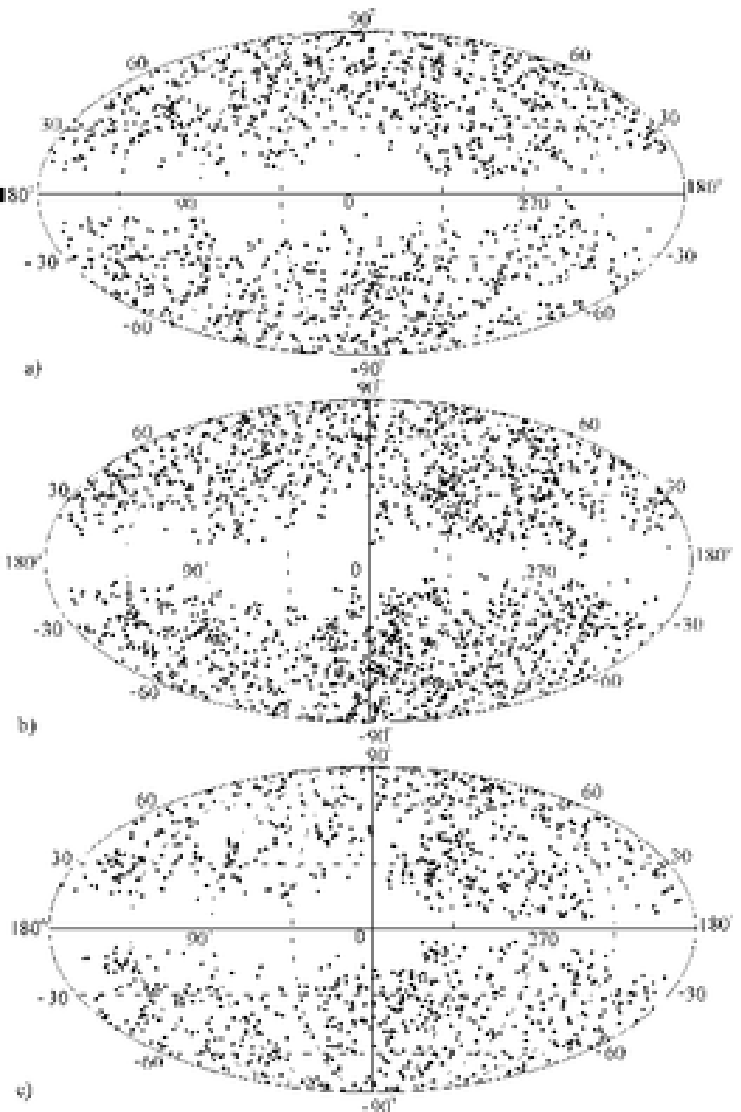,width=13cm}}
\caption{The distribution of the RFGC galaxies over the sky in galactic coordinates:
\newline a) $7<a/b<8$, b) $8<a/b<10$, c) $a/b>10$.}
\end{figure*}

\clearpage

\setcounter{page}{16}

{\small

\begin{table*}
\begin{center}
\caption{RFGC galaxies with different axial ratios  and angular
diameters.}

\vspace{0.3cm}

\begin{tabular}{|c|r|r|r|r|r|} \hline
&\multicolumn{4}{|c|}{Blue angular diameter, arcmin}& \\ \cline{2-5}
\multicolumn{1}{|c}{Axial ratio}&
\multicolumn{1}{|c}{$\geq$2.0}&
\multicolumn{1}{|c}{1.99-1.50}&
\multicolumn{1}{|c}{1.49-1.00}&
\multicolumn{1}{|c}{0.99-0.60}&
\multicolumn{1}{|c|}{All}\\ \hline
7.0 - 7.99& 95& 89& 275& 845& 1304 \\
8.0 - 9.99 & 124& 152& 407& 1116& 1799\\
$\geq$10.0&124& 143& 492& 374& 1133\\
\hline
All& 343& 384& 1174& 2335& 4236 \\ \hline
\end{tabular}
\end{center}
\end{table*}

\begin{table*}
\begin{center}
   \caption{Galaxies with different apparent magnitudes
    and angular
   diameters.}
\vspace{0.3cm}
   \begin{tabular}{|c|r|r|r|r|r|} \hline
   &\multicolumn{4}{|c|}{Angular diameter, arcmin}& \\ \cline{2-5}
   \multicolumn{1}{|c}{Magnitude, $B_t$}&
   \multicolumn{1}{|c}{$\geq$2.0}&
   \multicolumn{1}{|c}{1.99-1.50}&
   \multicolumn{1}{|c}{1.49-1.00}&
   \multicolumn{1}{|c}{0.99-0.60}&
   \multicolumn{1}{|c|}{All}\\ \hline
   $<13.0$   & 12&  0&   0&   0&   12 \\
   13.0 - 13.99&  32&   0&   0&    0&   32\\
   14.0 - 14.99&146&   1&   0&   0&  147\\
   15.0 - 15.99& 148 & 327& 216& 0& 691 \\
   16.0 - 16.99& 5& 56& 949& 1407& 2417\\
   17.0 - 17.99 & 0&0&9&928&937\\
   \hline
   All& 343& 384& 1174& 2335& 4236\\  \hline
   \end{tabular}
\end{center}
   \end{table*}

   \begin{table*}
\begin{center}
   \caption{Galaxies with different morphological types
    and angular
   diameters.}
\vspace{0.3cm}
   \begin{tabular}{|c|r|r|r|r|r|} \hline
   &\multicolumn{4}{|c|}{Angular diameter, arcmin}& \\ \cline{2-5}
   \multicolumn{1}{|c}{Type}&
   \multicolumn{1}{|c}{$\geq$2.0}&
   \multicolumn{1}{|c}{1.99-1.50}&
   \multicolumn{1}{|c}{1.49-1.00}&
   \multicolumn{1}{|c}{0.99-0.60}&
   \multicolumn{1}{|c|}{All}\\ \hline
  $Sab$       &  4&  0&   4&   2&   10 \\
$Sb$            &  23&  35&  46&   47&  151\\
$Sbc$           & 60&  76& 175& 262&  573\\
$Sc$            & 85&101 & 398& 952&   1536 \\
$Scd$           &72& 74& 254& 559 &  959\\
$Sd$             &65& 62& 214& 377& 718\\
$Sdm$ & 30& 27& 76& 119& 252\\
$Sm$& 4& 9& 7& 17& 37\\
   \hline
   All& 343& 384& 1174& 2335& 4236 \\ \hline
   \end{tabular}
\end{center}
   \end{table*}

   \begin{table*}
\begin{center}
   \caption{Disrtibution of galaxies according to the surface
brightness index for different
   diameters.}
\vspace{0.3cm}
   \begin{tabular}{|c|r|r|r|r|r|} \hline
   &\multicolumn{4}{|c|}{Angular diameter, arcmin}& \\ \cline{2-5}
   \multicolumn{1}{|c}{S.B.}&
   \multicolumn{1}{|c}{$\geq$2.0}&
   \multicolumn{1}{|c}{1.99-1.50}&
   \multicolumn{1}{|c}{1.49-1.00}&
   \multicolumn{1}{|c}{0.99-0.60}&
   \multicolumn{1}{|c|}{All}\\ \hline
     I     &56 & 39& 79& 68 &  242 \\
     II        &210 & 231& 671& 1368& 2480\\
    III        &63 & 105& 391& 810& 1369\\
    IV         & 14  &   9&  33&89& 145 \\
   \hline
   All& 343& 384& 1174& 2335& 4236 \\ \hline
   \end{tabular}
\end{center}
   \end{table*}

   \begin{table*}
\begin{center}
   \caption{Galaxies with different  index
   of
   asymmetry.}
\vspace{0.3cm}
   \begin{tabular}{|c|r|r|r|r|r|} \hline
   &\multicolumn{4}{|c|}{Angular diameter, arcmin}& \\ \cline{2-5}
   \multicolumn{1}{|c}{Asymmetry}&
   \multicolumn{1}{|c}{$\geq$2.0}&
   \multicolumn{1}{|c}{1.99-1.50}&
   \multicolumn{1}{|c}{1.49-1.00}&
   \multicolumn{1}{|c}{0.99-0.60}&
   \multicolumn{1}{|c|}{All}\\ \hline
	   0 & 231&248  & 740&1611 & 2830 \\
	    1   &  89& 103& 353&  614& 1159\\
	      2&23 & 33 & 81 & 110&  247\\
   \hline
   All& 343& 384& 1174& 2335& 4236 \\ \hline
   \end{tabular}
\end{center}
   \end{table*}

   \begin{table*}
\begin{center}
   \caption{Galaxies with different number of
   significant companions.}
\vspace{0.3cm}
   \begin{tabular}{|c|r|r|r|r|r|} \hline
  \multicolumn{1}{|c|}{Number of} &\multicolumn{4}{|c|}{Angular diameter, arcmin}& \\ \cline{2-5}
   \multicolumn{1}{|c}{compan.}&
   \multicolumn{1}{|c}{$\geq$2.0}&
   \multicolumn{1}{|c}{1.99-1.50}&
   \multicolumn{1}{|c}{1.49-1.00}&
   \multicolumn{1}{|c}{0.99-0.60}&
   \multicolumn{1}{|c|}{All}\\ \hline
     0& 205&   209& 539& 966& 1919 \\
	      1&  78&  99& 326&  675& 1178\\
	      2& 37&  37& 167& 377&  618\\
	      3&  11 &  22&  84&185& 302 \\
     $\geq4$   &12& 17&  58&  132&  219\\
   \hline
   All& 343& 384& 1174& 2335& 4236\\  \hline
   \end{tabular}
\end{center}
   \end{table*}

   \begin{table*}
\begin{center}
   \caption{Galaxies with different apparent magnitudes
    and axial
   ratios.}
\vspace{0.3cm}
   \begin{tabular}{|c|r|r|r|r|} \hline
   &\multicolumn{3}{|c|}{Axial ratio}& \\ \cline{2-4}
   \multicolumn{1}{|c}{Magnitude, $B_t$}&
   \multicolumn{1}{|c}{7.0-7.99}&
   \multicolumn{1}{|c}{8.0-8.99}&
   \multicolumn{1}{|c}{$\geq10.0$}&
   \multicolumn{1}{|c|}{All}\\ \hline
   $<13.0$   & 4 &  7&     1&   12 \\
   13.0 - 13.99&  15&  15&   2   &   32\\
   14.0 - 14.99& 61&  47&     39&  147\\
   15.0 - 15.99& 212 & 310& 169&  691 \\
   16.0 - 16.99&707&1008& 702 &2417\\
   17.0 - 17.99 &305& 412& 220& 937\\
   \hline
   All& 1304& 1799& 1133&     4236 \\ \hline
   \end{tabular}
\end{center}
   \end{table*}

   \begin{table*}
\begin{center}
   \caption{Galaxies with different morphological types
    and axial
   ratios.}
\vspace{0.3cm}
   \begin{tabular}{|c|r|r|r|r|} \hline
   &\multicolumn{3}{|c|}{Axial ratio}& \\ \cline{2-4}
   \multicolumn{1}{|c}{Type}&
   \multicolumn{1}{|c}{7.0-7.99}&
   \multicolumn{1}{|c}{8.0-8.99}&
   \multicolumn{1}{|c}{$\geq10.0$}&
   \multicolumn{1}{|c|}{All}\\ \hline
   $Sab$  & 5 &  4&     1&   10 \\
   $Sb$        &  77&  55&  19   &  151\\
   $Sbc$       & 240&279&     54&  573\\
   $Sc$        & 477 & 649& 410& 1536 \\
   $Scd$       &253& 438& 268 & 959\\
   $Sd$         &129& 265& 324& 718\\
   $Sdm$ & 98& 101& 53& 252\\
   $Sm$  & 25& 8& 4& 37\\
   \hline
   All& 1304& 1799& 1133&     4236\\  \hline
   \end{tabular}
\end{center}
   \end{table*}

   \begin{table*}
\begin{center}
   \caption{Galaxies with different surface brightness
  index     and axial
   ratio.}
\vspace{0.3cm}
   \begin{tabular}{|c|r|r|r|r|} \hline
   &\multicolumn{3}{|c|}{Axial ratio}& \\ \cline{2-4}
   \multicolumn{1}{|c}{S.B.}&
   \multicolumn{1}{|c}{7.0-7.99}&
   \multicolumn{1}{|c}{8.0-8.99}&
   \multicolumn{1}{|c}{$>10.0$}&
   \multicolumn{1}{|c|}{All}\\ \hline
      I   &99 &114&    29&  242 \\
     II        & 750&1123& 607   & 2480\\
     III       & 407&514&    448& 1369\\
     IV        & 48  &  48&  49&  145 \\
   \hline
   All& 1304& 1799& 1133&     4236 \\ \hline
   \end{tabular}
\end{center}
   \end{table*}

   \begin{table*}
\begin{center}
   \caption{Galaxies with different
  index of asymmetry    and axial
   ratio.}
\vspace{0.3cm}
   \begin{tabular}{|c|r|r|r|r|} \hline
   &\multicolumn{3}{|c|}{Axial ratio}& \\ \cline{2-4}
   \multicolumn{1}{|c}{Asymm.}&
   \multicolumn{1}{|c}{7.0-7.99}&
   \multicolumn{1}{|c}{8.0-8.99}&
   \multicolumn{1}{|c}{$\geq10.0$}&
   \multicolumn{1}{|c|}{All}\\ \hline
      0   &806&1210&   814& 2835 \\
      1        & 406& 490& 263   & 1154\\
      2        &  92& 99&     56&  247\\
   \hline
   All& 1304& 1799& 1133&     4236 \\ \hline
   \end{tabular}
\end{center}
   \end{table*}

  \begin{table*}
\begin{center}
  \caption{Galaxies with different number
of significant
  companions.}
\vspace{0.3cm}
  \begin{tabular}{|c|r|r|r|r|} \hline
  \multicolumn{1}{|c|}{Number of}&\multicolumn{3}{|c|}{Axial ratio}& \\ \cline{2-4}
  \multicolumn{1}{|c}{ compan.}&
  \multicolumn{1}{|c}{7.0-7.99}&
  \multicolumn{1}{|c}{8.0-8.99}&
  \multicolumn{1}{|c}{$\geq10.0$}&
  \multicolumn{1}{|c|}{All}\\ \hline
     0   &557&787&   575& 1919 \\
     1        & 394& 486& 298   & 1178\\
     2        & 192&289&    137&  618\\
     3        & 92 & 135& 75& 302\\
    $\geq4$   & 69  & 102&  48&  219 \\
  \hline
  All& 1304& 1799& 1133&     4236 \\ \hline
  \end{tabular}
\end{center}
  \end{table*}

\begin{table*}
\begin{center}
\caption{Centroid position in galactic coordinates for RFGC galaxies
with different angular diameters.}
\vspace{0.3cm}
\begin{tabular}{|c|r|r|r|r|r|} \hline
\multicolumn{1}{|c}{Angular diameters}&
\multicolumn{1}{|c}{$X$}&
\multicolumn{1}{|c}{$Y$}&
\multicolumn{1}{|c}{$Z$}&
\multicolumn{1}{|c}{$l\degr$}&
\multicolumn{1}{|c|}{$b\degr$}\\
\multicolumn{1}{|c|}{arcmin}  & & & & &\\
\hline
$\geq2.0$& $-0.076\pm0.027$& $-0.077\pm0.029$& $0.136\pm0.036$&225&$+15$\\
1.99 - 1.50& $-0.050\pm0.027$& $-0.019\pm0.028$& $0.113\pm0.032$&201&$+65$\\
1.49 - 1.00& $-0.050\pm0.015$& $-0.014\pm0.016$& $0.073\pm0.019$&196&$+54$\\
0.99 - 0.60& $0.052\pm0.010$& $-0.101\pm0.011$& $-0.033\pm0.014$&297&$-16$\\
\hline
All&      $0.004\pm0.008$& $-0.067\pm0.008$& $0.023\pm0.010$&273&$+19$\\
\hline
\end{tabular}
\end{center}
\end{table*}

}





\end{document}